\documentclass[12pt,preprint]{aastex}
\usepackage{apjfonts}

\newcommand{\lya}{\mbox{\,Ly$\alpha$}}

\newcommand{\mnii}{\ion{Mn}{2}}
\newcommand{\feii}{\ion{Fe}{2}}
\newcommand{\mgii}{\ion{Mg}{2}}
\newcommand{\mgi}{\ion{Mg}{1}}
\newcommand{\crii}{\ion{Cr}{2}}
\newcommand{\znii}{\ion{Zn}{2}}
\newcommand{\hi}{\ion{H}{1}}

\begin{document}

\title{GALEX Discovery of a Damped \lya\ System at Redshift z $\approx$1
\altaffilmark{1,2}}
\author{Eric M. Monier}
\affil{Department of Physics, The College at Brockport, 
State University of New York, \\ Brockport, NY 14420}
\email{emonier@brockport.edu}
\and
\author{David A. Turnshek, Sandhya M. Rao, and Anja Weyant}
\affil{Department of Physics \& Astronomy, University of 
Pittsburgh, Pittsburgh, PA 15260}

\medskip

\altaffiltext{1}{Based on observations made with the NASA Galaxy Evolution Explorer.  
GALEX is operated for NASA by the California Institute of Technology under NASA contract NAS5-98034.}
\altaffiltext{2}{Based in part on data obtained from the Sloan Digital Sky Survey.}




\begin{abstract}

We report the first discovery of a QSO damped \lya\ (DLA) system by
the $GALEX$ satellite. The system was initially identified as a
\mgii\ absorption-line system ($z_{abs}=1.028$) in the spectrum of 
SDSS QSO J0203$-$0910 ($z_{em}=1.58$).  
The presence of unusually strong absorption due to metal lines of
\znii, \crii, \mnii, and \feii\ clearly suggested that it might be a 
DLA system with $N_{HI} \ge 2 \times 10^{20}$ atoms cm$^{-2}$.
Follow-up $GALEX$ NUV grism spectroscopy confirms the 
system exhibits a DLA absorption line, with a
measured \hi\ column density of $N_{HI} = 1.50\pm0.45 \times 10^{21}$ atoms cm$^{-2}$.    
By combining the $GALEX$ $N_{HI}$ determination with the SDSS spectrum measurements of 
unsaturated metal-line absorption due to \znii, which is generally not depleted onto grains, 
we find that the system's neutral-gas-phase 
metal abundance  is [Zn/H] = $-0.69\pm0.22$, or $\approx$20\% solar. 
By way of comparison, although this system has one of the largest Zn$^+$ column
densities, its metal abundances are comparable to 
other DLAs at $z\approx1$.
Measurements of the abundances of Cr, Fe, and Mn help to further pin
down the evolutionary state of the absorber.

\end{abstract}


\section{Introduction}

Since the first survey for damped \lya\ systems (DLAs) by \cite{wolf86},
it has been recognized that these galaxy-sized columns of neutral hydrogen 
($N_{HI} \ge 2 \times 10^{20}$ atoms cm$^{-2}$) 
trace the neutral gas content of the universe back 
to the epoch of the earliest QSOs.  Spectroscopic surveys for DLAs rely
on only the presence of a background QSO, and therefore probe the universe 
independently of normal galaxy imaging.  As such, they do not depend on
the luminosity or spectral energy distribution of any associated galaxies. 
Only the presence of significant dust in the absorber would cause a bias 
against the identification of a DLA system. 
Surveys have shown that DLAs trace the bulk of the 
observable \hi\ gas mass in the universe 
\citep[][hereafter RTN2006]{procha05, rtn06}.
\defcitealias{rtn06}{RTN2006}

Several facets of galaxy formation and evolution can be considered through 
follow-up studies of DLA properties.   Their metal abundances, for example,
provide insight into the chemical evolution of the neutral gas over cosmic timescales.
Follow-up studies at $z>1.65$ generally rely on blind spectroscopic surveys for \lya, for which the \lya\  line 
is shifted to optical wavelengths accessible from the ground (e.g. Prochaska et al. 2005 and references therein).  
However, these high-redshift DLAs
only provide insight into the cosmic chemical evolution of neutral gas during
the first third of the age of the universe \citep[e.g.][and references
therein]{pet02, procha03, procha07}.

At redshifts $z<1.65$, Hubble Space Telescope ($HST$)-targeted searches for DLAs in strong MgII 
systems, like those found in SDSS spectra by \cite{nes05},
have resulted in significant numbers of lower-redshift DLAs being identified
(e.g. RTN2006 and references therein). This has led to a growing sample of DLAs that have been used to determine 
cosmic neutral gas-phase metallicities at lower redshifts \citep[e.g.][and references therein]{rao05, nest08}. 
However, the failure of the Space Telescope Imaging Spectrograph (STIS) onboard $HST$ in
2004 has curtailed further lower-redshift DLA surveys. Moreover, lacking $HST$ UV spectroscopy, follow-up metallicity studies of the 
lowest redshift population of DLAs have become impossible, given that the weak unsaturated 
metal lines that must be observed (\ion{Cr}{2}, \ion{Fe}{2}, \ion{Mn}{2}, \ion{Si}{2}, 
\ion{Zn}{2}) lie in the UV when the redshifts are $z<0.6$.  
Therefore, until the Cosmic Origins Spectrograph (COS) becomes available onboard $HST$, the 
best opportunities for measuring individual DLA metallicities 
at $z<1.65$ lie at redshifts $z>0.6$, for which the \ion{Zn}{2} $\lambda\lambda$2026,2062 
lines are shifted into the optical regime. In total there are currently only about 30
DLA systems with $0.6<z<1.65$ that can be used for individual metallicity studies.

Alternatively, an additional possibility is that DLA systems in the redshift interval $0.6<z<1.65$ 
might be identified by studying optical spectra and selecting systems 
which exhibit unusually strong metal-line absorption due to, e.g., \znii. 
Such systems are representative of the highest metal-line 
column densities in the universe, and are likely to have \hi\ column 
densities in the DLA regime.

In this contribution, we successfully employ this second approach. After identifying a strong metal-line
system at $z_{abs} \approx 1$ in the spectrum of SDSS QSO J0203$-$0910, we obtained 
follow-up NUV grism spectroscopy from the NASA Galaxy Evolution Explorer ($GALEX$) satellite 
to confirm our hypothesis that it is a DLA system. 
In particular, the damping wings on the Voigt profile of the observed \lya\ line 
are strong and broad 
enough that $N_{HI}$ is measurable even in the low-resolution $GALEX$ NUV 
grism spectrum.
This represents the first discovery of a DLA system by $GALEX$. 
Moreover, given that the metal lines of the absorption system are visible and measurable 
in the SDSS spectrum, our results also lead to a metallicity determination. This adds to the 
growing number of DLA metallicity 
measurements at $z < 1.65$.   

The paper is laid out as follows.  In \S2, we present the $GALEX$ data
and our measurement of $N_{HI}$; we also present the SDSS spectrum
of QSO J0203$-$0910.  We describe our measurements of the metal-line 
equivalent widths in \S3.  In \S4 we derive and present the metal 
abundances of this system.  We summarize 
our results in \S5 and compare them to previous studies.

\section{Observations}

The SDSS QSO J0203$-$0910 was discovered in SDSS DR1 \citep{sch03, rich04}. 
The QSO has SDSS u, g, r, i, and z magnitudes of 18.6, 18.3, 18.2, 18.0, and 18.0, respectively. 
This $z_{em}=1.58$ QSO has a strong \mgii\ system
along the sightline at $z_{abs} = 1.028$, with a rest equivalent width of 
$W_0^{\lambda2796}=2.66 \pm 0.06$ \AA.\footnote{\cite{procht06} report 
$z_{abs}=1.027$ and $W_0^{\lambda2796} = 2.36$ \AA.}
Since this line is saturated, the equivalent width is in fact an indication
of the absorber's velocity spread, or $\Delta$v $\approx$ 285 km s$^{-1}$.

We selected this system as a candidate DLA based on the relatively 
strong \znii, \crii, \mnii, and (some) \feii\ absorption lines present in the SDSS spectrum 
(\S2.2).  It was approved for 31,500 sec of observation during Cycle 3 
of $GALEX$ (Program ID 0104).

\subsection{$GALEX$ Data}

\begin{figure}
\plotone{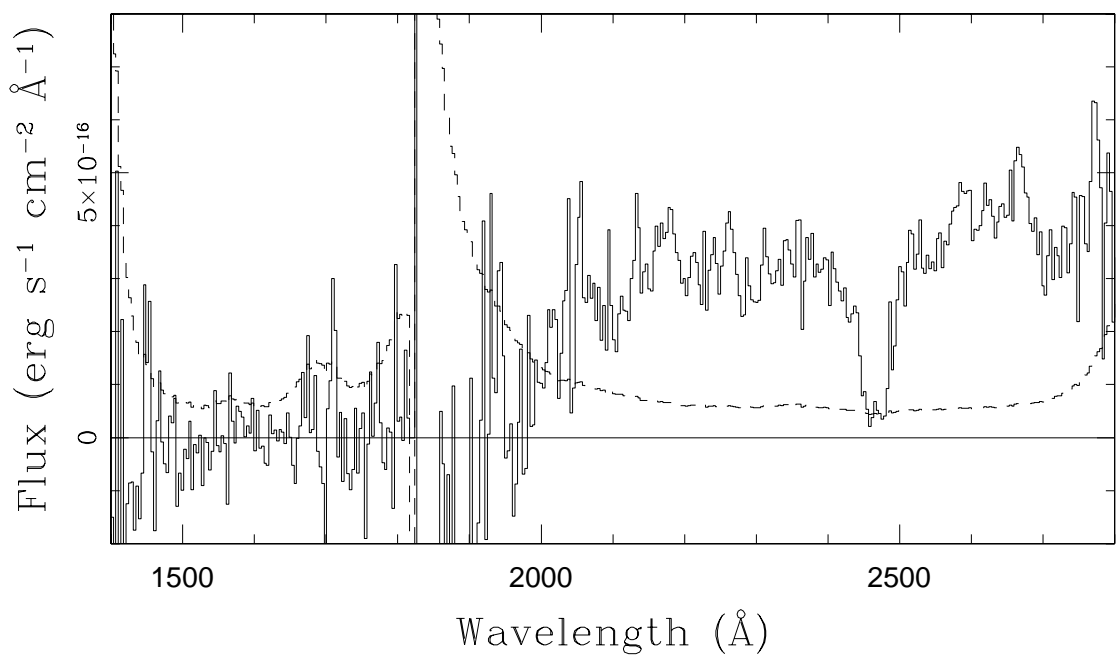}
\caption[]{The complete $GALEX$ FUV-NUV grism spectrum 
of SDSS QSO J0203$-$0910.  The $z_{abs} = 1.028$ \lya\ absorption 
line is clearly visible near 2465 \AA.  The dashed line is the 1-$\sigma$
error array.  The FUV and NUV spectra meet at 1750 \AA, where the 
detectors have minimal sensitivities.  The drop in sensitivity 
of the detectors and the Lyman limit of this absorber 
(at $\approx$ 1850 \AA) are responsible for the spike in the error
array and the absence of flux in the FUV channel.
\label{fullspec}}
\end{figure}

The $GALEX$ grism spectra of SDSS QSO J0203$-$0910 ($GALEX$ NUV mag = 19.8) 
were obtained in November 2007. The total exposure time of 30,720 sec was split between 16 visits.  These data
were processed, combined, and extracted in the standard pipeline procedure 
described by \cite{mor07}.   The resulting combined FUV-NUV spectrum is shown in 
Figure \ref{fullspec}.  
The spectrum has a signal-to-noise ratio of $S/N \approx 6$ per pixel in the region of the DLA
at $\approx$ 2465 \AA, where the resolution of the spectrum is $R \approx 117$ with 6 pixels
per resolution element. 

\begin{figure}
\plotone{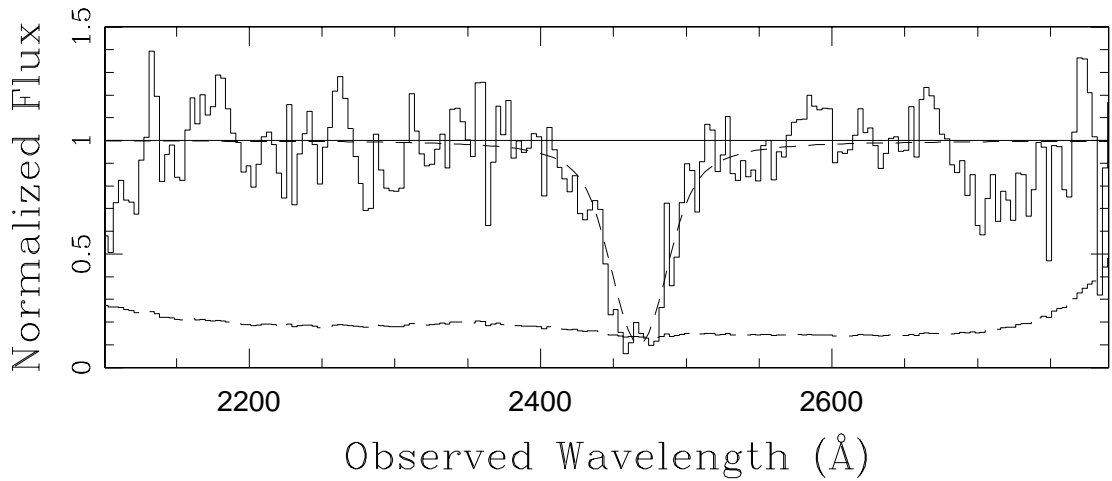}
\caption[]{The normalized observed-frame $GALEX$ NUV grism spectrum 
of SDSS QSO J0203$-$0910 and the \lya\ absorption line at
$z_{abs} = 1.028$.  The line has 
been fitted with a Voigt profile having column density 
$N_{HI}= 1.5\times$10$^{21}$ atoms cm$^{-2}$ and convolved with the 
$GALEX$ NUV grism resolution.
The horizontal dashed curve is the 1$\sigma$ error array.
\label{galexspec}}
\end{figure}

In Figure \ref{galexspec} we show the normalized $GALEX$ spectrum in the region of the
 $z_{abs} =1.028$ \lya\ absorption line.  
We fitted a local continuum to the spectral
region using standard IRAF\footnote{$IRAF$ is distributed by the National Optical Astronomy Observatory, 
which is operated by the Association of Universities for Research in Astronomy (AURA) 
under cooperative agreement with the National Science Foundation.} routines.  Since the absorption line lies in 
the \lya\ forest, using a least-squares minimization to fit the line with a Voigt profile is not the best way to proceed.  
Instead, the line was fitted by eye with a Voigt profile 
convolved to the instrumental resolution following the procedure described 
in \cite{rt00} and \citetalias{rtn06}.  The placement of the continuum in the 
region of the absorption line is the primary source of $N_{HI}$ uncertainty.  
To estimate the errors associated with the $N_{HI}$ measurement, 
we shifted our estimate of the best-fit continuum level upward and downward by 
the 1$\sigma$ error array and refit the DLA, as described in \cite{rt00}.  
The resulting column density determination is 
$N_{HI}=1.5^{+0.4}_{-0.5}\times10^{21}$ atoms cm$^{-2}$, which
we report as $1.5\pm0.45\times10^{21}$ atoms cm$^{-2}$.

\subsection{The SDSS QSO J0203$-$0910 Spectrum}

\begin{figure} 
\plottwo{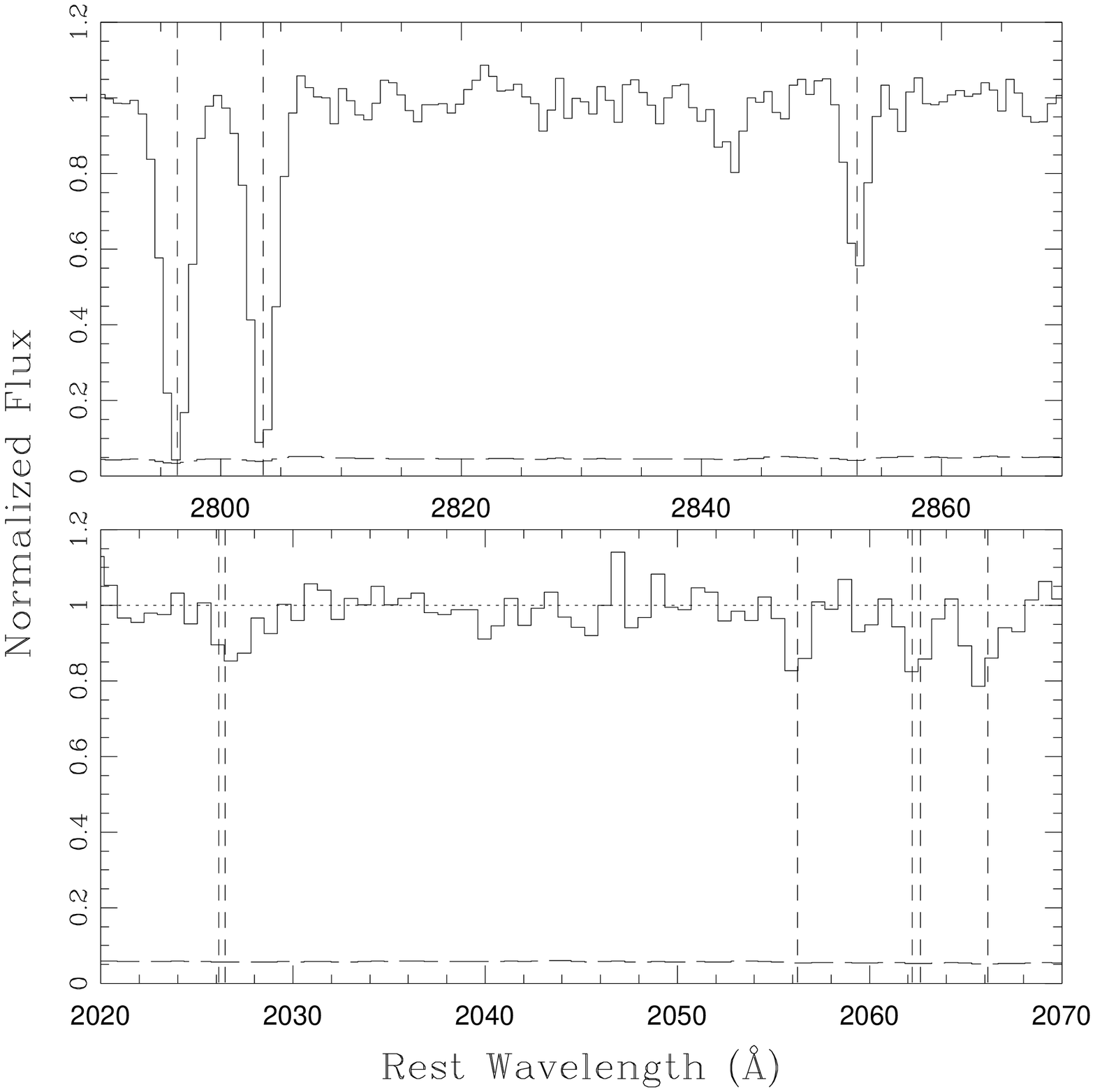}{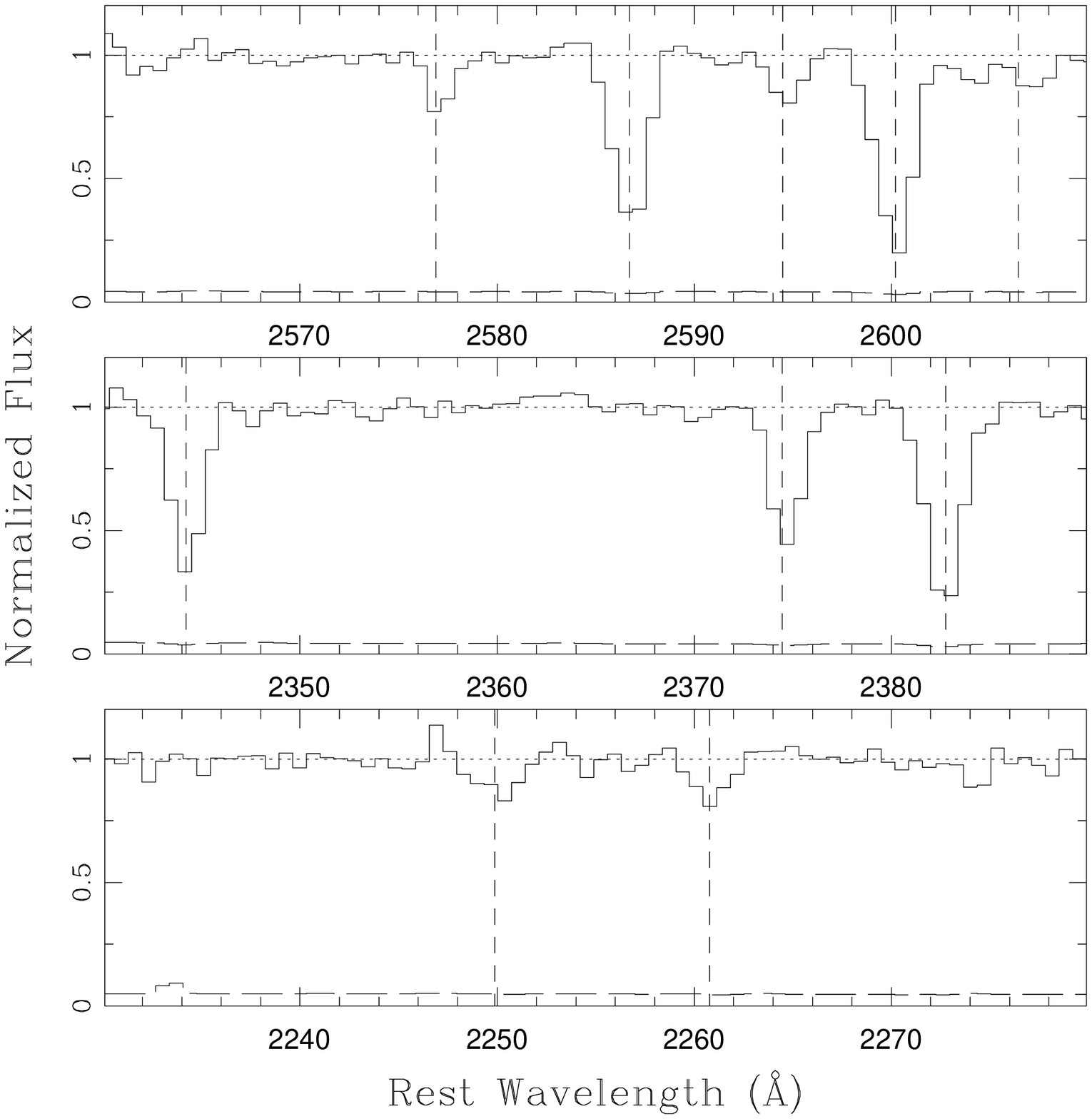}
\caption{These portions of the normalized spectrum of SDSS QSO J$0203-0910$ show 
selected absorption lines due to metals in the rest frame of the $z=1.028$ absorber. 
See Table 1 for the rest wavelengths of the various absorption lines marked by 
vertical dashed lines.   The spectra are unsmoothed, and the horizontal dashed 
curves show the 1$\sigma$ error array.  The two panels on the left side show the 
regions containing the strong \mgii\ doublet and the \mgi\ line (top panel) and 
the \znii, \mgi, and \crii\ lines (bottom panel).  The three panels on the right 
side show the regions containing \mnii\ and strong \feii\ lines (top panel), 
other strong \feii\ lines (middle panel), and weak \feii\ lines (bottom panel).
\label{sdssspec}}
\end{figure}

The SDSS spectrum of QSO J0203$-$0910 was fitted using a cubic spline for the 
continuum and Gaussians for the broad emission features, as described in \cite{nes05}.   
For presentation, portions of the spectrum revealing the relevant metal lines are shifted into the rest frame of the absorber and normalized by the 
continuum fit; they are shown in the five panels of Figure \ref{sdssspec}. The figure caption notes which metal-line 
regions are covered.

\section{Metal Absorption-Line Measurements}

The rest equivalent width measurements of various unsaturated metal absorption 
lines (e.g. \znii, \crii, \mnii, and \feii) in the $z_{abs}=1.028$ system are presented in Table 1 (column 3). 
For completeness, Table 1 also includes 
measurements of saturated lines due to  \mgii\ and \feii.  
Note that the \feii$\lambda$2600 rest equivalent 
width, in combination with the \mgii$\lambda$2796 rest equivalent width,
makes the absorber a DLA candidate system according to the criterion of 
\citetalias{rtn06}.
Details of measuring lines due to \znii, \crii, \mnii, and
\feii\ are discussed below since the results are used 
to determine metal abundances in \S4.

\subsection{\znii\ and \crii}

We measured the equivalent widths of the metal lines in the normalized 
SDSS spectrum by fitting Gaussians to the absorption features using the ratios
of relevant transition oscillator strengths, as presented by \cite{nes03} 
(see their Table 2).  The blends of \znii\ and \mgi\ at 2026 \AA\footnote{A \crii\ line is 
also part of this blend, but its oscillator strength is so small that it was ignored 
\citep{nes03}.} and 
\crii\ and \znii\ at 2062 \AA\ are unresolved and were fitted by single Gaussians.  
The rest equivalent width of \znii$\lambda$2026 was found by subtracting 
the contribution
due to \mgi$\lambda$2026, as inferred
from the measurement of the stronger \mgi$\lambda$2852 line. 
The rest equivalent width of \crii$\lambda$2062 was 
inferred from the fit to \crii$\lambda$2056; it was subtracted from the 
measurement of the \crii\ and \znii\  2062 \AA\ blend, resulting 
in the reported  \znii$\lambda$2062 measurement.  
We note that the relatively large measured rest equivalent width 
of \crii$\lambda$2066 (oscillator strength $f=0.0515$) seems 
anomalous given that it should be weaker than \crii$\lambda2056$ and \crii$\lambda2062$. 
Therefore, we have not used the \crii$\lambda$2066 
measurement for the Cr$^+$ column density and abundance 
determinations reported in \S4.

\subsection{\mnii\ and \feii}

Measurements of some of the \ion{Mn}{2} absorption lines at $z=1.028$ are slightly
complicated by \feii\ absorption lines in a weaker \mgii\ system 
($W_0^{\lambda2796}=0.442\pm0.057$ \AA) at $z_{abs} = 1.217$.
The \mnii$\lambda2576$ line ($z=1.028$) is unaffected, however 
\mnii$\lambda2594$ ($z=1.028$) is blended with \feii$\lambda2374$ 
($z = 1.217$) and \mnii$\lambda2606$ ($z=1.028$) falls near  
\feii$\lambda2382$ ($z=1.217$).

After measuring the absorption feature at 
the position of the \mnii$\lambda2594$ ($z=1.028$) and \feii$\lambda2374$ 
($z = 1.217$) blend, we evaluated the 
contribution from \feii\ as follows.  We first measured the \feii\ $\lambda2344$ and \feii\ 
$\lambda2600$ lines in the $z=1.217$ system, and used the relevant oscillator
stengths to infer the \feii\ $\lambda2374$ ($z=1.217$) rest equivalent width; its  
contribution is estimated to be $50\pm15$ m\AA.
We then subtracted this value from the total measured rest equivalent width of 
the \mnii\ and \feii\ blend to obtain $W_0^{\lambda2594} = 335
\pm 58$ m\AA\ ($z=1.028$).

The \mnii$\lambda$2606 line has the lowest oscillator strength of the three \mnii\ lines, $f=0.1927$.
Our measured value of this line, $W_0^{\lambda2606}=175$ $\pm$ 44 m\AA\ ($z=1.028$),
is in agreement with the value predicted from the 
isolated \mnii$\lambda2576$ line, $W_0^{\lambda2606}=185$ m\AA.
Also, the \feii$\lambda2382$
line ($z=1.217$) and the \mnii$\lambda2606$ line ($z=1.028$)
should be approximately resolved, given their $\approx$ 4 \AA\ separation
in the observed frame.  We conclude, therefore, that the
contribution of \feii\ $\lambda2382$ to the \mnii\ $\lambda2606$
measurement is negligible.  

There are five \feii\ lines that can be measured in the SDSS spectrum,
although only two of them, \feii$\lambda2249$ and \feii$\lambda2260$,
appear to have equivalent widths placing them on the linear
part of the curve of growth.  Thus, these are the two most useful \feii\
lines for column density determinations. It should
be noted that the $z=1.217$ \crii\ lines at 2056 \AA\ and 2066 \AA\  
are potential contaminants of the $z=1.028$ \feii\  lines at 2249 \AA\ and 
2260 \AA, respectively. However, inspection of the SDSS spectrum 
indicates no evidence for  \znii$\lambda2026$ absorption or a \znii$\lambda2062$ and \crii$\lambda2062$ 
absorption blend at $z=1.217$.  Given this, we conclude that 
any \znii\ or \crii\ features at $z=1.217$ are too weak to produce a significant
detection in the SDSS spectrum; they would certainly lie well within the 
reported measurement errors 
for the \feii$\lambda2249$ and \feii$\lambda2260$ lines at $z=1.028$.

\section{Column Densities and Element Abundance Determinations}

With the exception of the \crii$\lambda2066$ measurement described above, 
but which we ignore, the other rest equivalent width measurements appear 
to be in reasonable agreement.
The rest equivalent widths of the weak lines due to 
\znii, \crii, \mnii, and two of the \feii\ lines 
can be converted directly into column densities since they
are unsaturated and lie on the linear part of the curve
of growth.

Column density results for these ions are presented in Table 1 (column 4) next to the
first transition of each ion; the results are weighted averages derived from all
of the relevant rest equivalent widths in column 3, unless otherwise noted.
Table 1 also lists the $N_{HI}$ value determined 
from the $GALEX$ spectrum.  The various column densities are used to derive the
element abundance results discussed below.  When we quote abundances relative 
to solar values, we use the solar values of \cite{gs98} as compiled in Table 2 of 
\cite{rao05}.  

It is generally agreed that most of the Zn and Cr is singly 
ionized in the neutral DLA regions \citep[]{hs99}; this should hold for
singly ionized Fe and Mn as well.
Under this assumption, we have converted the Zn$^+$, Cr$^+$, Mn$^+$, and Fe$^+$
column densities into the neutral gas-phase abundances.
We find that [Zn$/$H] = $-0.69\pm0.22$, or $\approx$20\% solar;  
[Cr$/$H] = $-1.07\pm0.18$ and [Fe$/$H] = $-1.03\pm0.16$, both
$\approx$9\% solar;  and [Mn$/$H] = $-1.48\pm0.17$, or $\approx$3\% solar.  These 
results are given in Table 1 (column 5).

Also in Table 1 (column 6), we give the elemental abundances 
of Zn, Cr, and Mn relative to Fe, expressed as [X/Fe].  
We find [Zn/Fe] = $0.34\pm0.15$ and [Cr/Fe] = $-0.04\pm0.07$.
The value [Cr/Zn] = $-0.38\pm$0.17 is typically taken 
as an indication of the presence of dust, given that Cr is expected to be
readily depleted onto grains, whereas Zn is not \citep{pet90}.
The same holds for the inferred value of [Fe/Zn].
Thus, from the measurements we can infer 
that $\approx$60\% of the Cr (and Fe) is depleted onto grains.  
This value is comparable to [Cr$/$Zn] $\approx -0.5$ found for
the Milky Way halo \citep{sav96}.
\cite{nes03} found  [Cr$/$Zn] = $-0.64\pm0.13$ in a composite SDSS 
spectrum for a sample of \mgii\ absorbers with $W_0^{2796} \ge 1.3$\AA\ 
and $0.9\le z\le1.3$. 
With regard to Mn, we also find that [Mn/Fe] = $-0.45\pm0.16$.  This 
result is similar to that found in previous DLA metallicity studies 
\citep[e.g.,][]{pet00, led02, rao05}; it has been suggested to be 
due to metallicity-dependent production of Mn
(see \S 5).

\section{Summary \& Discussion}

Based on the presence of unusually strong metal lines (\znii, \crii, \mnii, and two 
low-oscillator-strength \feii\ lines) at $z_{abs}=1.028$ in the SDSS spectrum of 
QSO J0203$-$0910, we selected the  system as a DLA absorber candidate. 
We obtained a $GALEX$ NUV grism spectrum to confirm the presence
of the DLA, and found $N_{HI}=1.5^{+0.4}_{-0.5}\times10^{21}$ atoms cm$^{-2}$. 
This marks the first discovery of a DLA system by $GALEX$.
Using the existing SDSS spectrum,  we determined the system's metal-line 
column densities from measurements of the unsaturated metal lines. 

As discussed further below, the Zn$^+$ column density in this system is among the highest
known. Therefore, prior to our $GALEX$ observation, we also thought 
that this system might be a good candidate for a solar or super-solar metallicity
DLA system. However, we found  [Zn$/$H] = $-0.69\pm0.22$, or $\approx$20\% solar, 
which is within the normal range for DLA systems at this redshift.

Measurements of absorption
lines due to \crii\ and \feii\  in this system indicate that the gas-phase abundance of Cr and Fe is lower
than Zn, with $\approx 60\%$ of the Cr and Fe 
depleted onto grains. In addition, measurements of the \mnii\ lines indicate that, relative to solar, the 
gas-phase abundance of Mn is even less than that of Cr and Fe. 
This may be due to a combination of depletion and an early stage of chemical evolution 
for the absorbing gas.  In particular, it is agreed that [Mn$/$Fe] is clearly underabundant
(e.g. [Mn$/$Fe] $< -0.2$) in low-metallicity stars (e.g. [Fe$/$H] $<$ -1), but then
[Mn$/$Fe] starts to increase with increasing [Fe$/$H] \citep[][and references therein]{mcw03}.
However, the details of the origin of Mn production, and the roles of SNeIa and SNeII 
enrichment, are still debated \citep[e.g.][and references therein]{ffb07, bg08}.

Our value for the Zn$^+$ column density,
log N$_{Zn^+} \approx 13.16\pm0.15$, places it right at the empirical 
threshold of log N$_{Zn^+} < 13.15$ on a plot 
of [Zn/H] vs. log $N_{HI}$, above which DLAs have generally not been 
found \citep{boi98, mei06}.  \cite{hf06} have classified DLAs above 
this threshold, i.e.,  with log N$_{Zn^+} \ge 13.15$, as ``metal-strong.''
This soft limit has been suggested as being due to obscuration 
by dust \citep[e.g.][]{boi98}, however more recent results suggest that 
the lack of systems above this limit is due more to their inherent rarity  
\citep[and references therein]{hf06}.

The strong metal lines in this system, nevertheless, 
are not indicative of an unusually large metallicity compared 
to other DLAs at 
$z\approx$1.  For example, \cite{kul07} found $<$[Zn/H]$>=-0.82\pm0.15$ 
for 20 DLAs with $0.1<z<1.2$, which is consistent with our result.  
Rather, the strong metal lines in this system can be attributed to its large 
\hi\ column density.  As noted above, this DLA does have a relatively high 
metallicity in comparison to systems with similar $N_{HI}$.  

Our metallicity results may also be compared to those determined
at $z>0.6$ from SDSS composite spectra.  For
example, in an analysis of one composite created from 90 \mgii\ 
absorbers with $W_0^{\lambda2796}> 1.3$ \AA\ and 
$0.9 < z < 1.35$, many of which should be DLAs 
\citetext{\citealp{rt00}; \citetalias{rtn06}},
\citet{nes03} estimated a metallicity of [Zn$/$H] =$ -0.56\pm0.19$.  
\cite{daveiau05} reported an extension of this analysis using a much large sample of 
strong \mgii\ absorbers with 1$\lesssim z \lesssim2$, and parameterized
[Zn$/$H] = $-0.4-0.0043e^{[(6-W_0^{\lambda2796})/0.88]}$ for 0.6\AA$\leq 
W_0^{\lambda 2796} \leq 6.0$\AA.  Using our measured value of 
$W_0^{\lambda 2796}=2.66$ \AA\ for the \mgii\ absorber corresponding 
to the DLA in SDSS QSO J0203$-$0910, this paramaterization would 
predict [Zn$/$H] = $-0.59$, in agreement with our measured value of 
[Zn$/$H] = $-0.69\pm0.22$.  

Candidate DLA selection based on strong lines of \znii\
remains the best chance for finding solar or 
super-solar metallicity DLAs at $z\approx1$. We suspect that 
such a system might exhibit a Zn$^+$ column density similar
to the $z_{abs}=1.028$ system, but have $N_{HI} \approx
2 \times 10^{20}$ atoms cm$^{-2}$, i.e.,  right at the DLA 
$N_{HI}$ lower threshold. The presence of other normally
weak lines may also provide a clue, but many of the other elements are 
subject to depletion. Larger surveys
using $HST$ COS will permit an evaluation of the success of this technique
at $0.6\leq z \leq1.65$.

\acknowledgements

We thank Daniel Nestor and Anna Quider for their contributions 
which helped produce the Pittsburgh SDSS \mgii\ QSO absorption-line
catalog.  E.M.M  gratefully acknowledges support 
from NASA/$GALEX$ Guest Investigator grant NNX 08AJ87G. 
D.A.T and S.M.M. acknowledge support from NASA/$GALEX$ 
Guest Investigator grant NNX 08AE20G.

Funding for the SDSS and SDSS-II has been provided by the Alfred P. Sloan Foundation, 
the Participating Institutions, the National Science Foundation, the U.S. Department 
of Energy, the National Aeronautics and Space Administration, the Japanese 
Monbukagakusho, the Max Planck Society, and the Higher Education Funding 
Council for England. The SDSS Web Site is http://www.sdss.org/.

The SDSS is managed by the Astrophysical Research Consortium for the Participating 
Institutions. The Participating Institutions are the American Museum of Natural 
History, Astrophysical Institute Potsdam, University of Basel, University of Cambridge, 
Case Western Reserve University, University of Chicago, Drexel University, Fermilab, 
the Institute for Advanced Study, the Japan Participation Group, Johns Hopkins 
University, the Joint Institute for Nuclear Astrophysics, the Kavli Institute for 
Particle Astrophysics and Cosmology, the Korean Scientist Group, the Chinese Academy 
of Sciences (LAMOST), Los Alamos National Laboratory, the Max-Planck-Institute for 
Astronomy (MPIA), the Max-Planck-Institute for Astrophysics (MPA), New Mexico State 
University, Ohio State University, University of Pittsburgh, University of Portsmouth, 
Princeton University, the United States Naval Observatory, and the University of Washington.

\clearpage


\begin{deluxetable}{lccccc} 

\tablewidth{0in}

\tablenum{1}
\tablecaption{Absorption Line Measurements: 
SDSS QSO J0203$-$0910, $z=1.028  $
\label{metaldata}}
\tablehead{\colhead{Ion} & 
 \colhead{$\lambda_{rest}$}  &  \colhead{$W_0$} & \colhead{log $N$} & \colhead{[X/H]}
 &  \colhead{[X/Fe]}  \\
 \colhead{}  &  \colhead{(\AA)} & \colhead{(m\AA)} & \colhead{(cm$^{-2}$)}  & &
 }
\startdata
 \ion{H}{1}     &  1215.67   &  \nodata              &  $21.18^{+0.11}_{-0.16}$  & \nodata & \nodata \\
 \ion{Zn}{2}    &  2026.14   &   $306\pm91$       &    $13.16\pm0.15$     &  $-0.69\pm0.22$  &  $0.34\pm0.15$  \\
                &  2062.66   &   $ 92\pm65$       &                       \\
 \ion{Cr}{2}    &  2056.25   &  $235\pm50$    &  $13.78\pm0.07$\tablenotemark{a} & $-1.07\pm0.18$  & $-0.04\pm0.07$  \\
                &  2062.23   &  $175\pm37$    &                       \\
                &  2066.16   &  $381\pm63$    &                       \\
 \ion{Fe}{2}    &  2249.88   &  $371\pm53$    &  $15.65\pm0.03$\tablenotemark{b} &  $-1.03\pm0.16$ & \nodata      \\
                &  2260.78   &  $396\pm44$    &                       \\
                &  2344.21   & $1287\pm35$    &                       \\
                &  2374.46   & $1108\pm34$          &                       \\
                &  2382.77   & $1791\pm33$          &                       \\
                &  2586.65   &  $1430\pm46$         &                       \\
                &  2600.17   &  $1780\pm45$         &                       \\
 \ion{Mn}{2}    &  2576.88   &  $337\pm44$          &  $13.23\pm0.05$    & $-1.48\pm0.17$ & $-0.45\pm0.16$    \\
                &  2594.50   &  $335\pm58$\tablenotemark{c}          &                       \\
                &  2606.46   &  $176\pm45$\tablenotemark             &                       \\
 \ion{Mg}{1}    &  2026.48   &  $27\pm2$             &                       \\
                &  2852.96   &  $890\pm56$          &                       \\
 \ion{Mg}{2}    &  2796.35   &  $2265\pm56$         &                       \\
                &  2803.53   &  $2450\pm62$         &                       \\

\enddata
\tablenotetext{a}{Calculated from \crii\ $\lambda2056$, $\lambda2062$}
\tablenotetext{b}{Calculated from \feii\ $\lambda2249$, $\lambda2260$}
\tablenotetext{c}{Blend with weak \ion{Fe}{2} $\lambda$2374 line at $z_{abs}=1.217$}
\end{deluxetable}

\end{document}